\documentclass{moriond}

\def\be{\begin{equation}}
\def\ee{\end{equation}}
\def\bea{\begin{eqnarray}}
\def\eea{\end{eqnarray}}

\newcommand{\Photo}{\includegraphics[height=35mm]{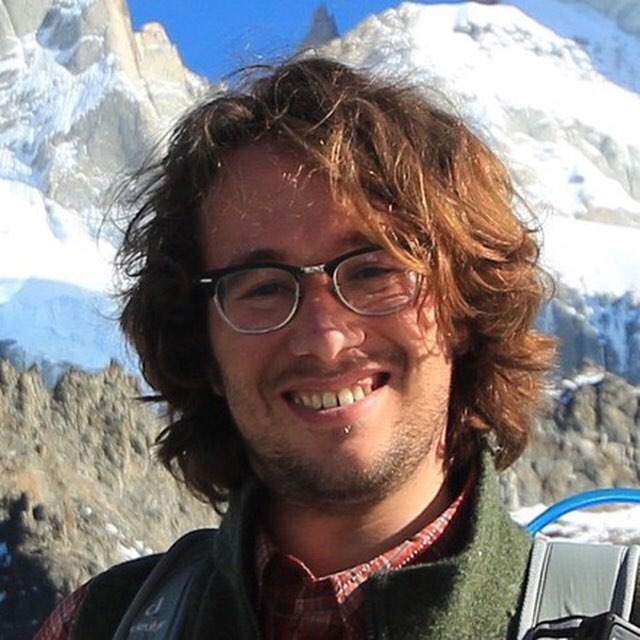}}

\begin{document}
\vspace*{4cm}
\title{HEAVY-QUARK PRODUCTION AT NNLO}

\author{SIMONE DEVOTO}

\address{Dipartimento di Fisica Aldo Pontremoli, Universit\`a di Milano and INFN, Sezione di Milano, Via Celoria 16, I-20133 Milano, Italy}

\maketitle\abstracts{
We present the calculation of QCD radiative corrections to heavy-quark production at NNLO.
The calculation is implemented in an updated version of \textsc{Matrix}, which allows us to obtain multidifferential predictions for stable top quarks.
We also report on the NNLO calculation for $b{\bar b}$ production and comment on the inclusion of top-quark decays and of the extension of our results to $t\bar tH$.}

\section{Introduction}
Top quark ($t$) production is a process of key importance for the LHC physics programme.
The mass of the top quark represents one of the fundamental parameters of the Standard Model: its high value leads to  a strong coupling with the Higgs boson and thus to an important role in electroweak-symmetry breaking.
Furthermore, the top quark is a possible window on new physics scenarios, and its production constitutes an important background for both precision Standard Model measurements and Beyond the Standard Model searches.
The main source of top events at the LHC is the production of a $t\bar t$ pair, whose measurement is now reaching an impressive experimental precision.
During the last decades, there have been many efforts from the theoretical community to provide predictions that can match such experimental accuracy.

In the following, we will focus on our computation of NNLO QCD corrections to top-quark pair production\cite{Catani:2019iny,Catani:2019hip,Catani:2020tko} that, through its implementation within the \textsc{Matrix}\cite{Grazzini:2017mhc} framework, provided a first public NNLO generator for the process.
We will briefly discuss further developments of this work, including bottom-quark (b) pair production\cite{Catani:2020kkl}, $t{\bar t}H$ production and $t{\bar t}$ production including top-quark decays.

\section{Top-quark pair production: inclusion in MATRIX 2.1.0}
Our computation of $t\bar t$ production relies on the $q_T$-subtraction formalism\cite{Catani:2007vq} as a method to handle and cancel infrared divergences.
Originally developed for the production of colourless final states, with the inclusion of extra contributions it can be extended to massive, colourful final states.

Within the $q_T$-subtraction formalism, the cross section for the production of a general final state $F$ can be written as follows:
\begin{equation}
  \label{eq:qt_sub}
  d\sigma^F_{NNLO}=\mathcal H^F_{NNLO}\otimes d\sigma_{LO}^F+\left[d\sigma_{NLO}^{F+{jet}}-d\sigma^{CT}_{NNLO}\right]\;.
\end{equation}
The term in the square brackets represents the contribution to the cross section with $q_T\neq 0$, $q_T$ being the transverse momentum of the system $F$.
Since the final state $F$ has $q_T=0$ at Born level kinematics, this contribution at NNLO is effectively captured by the NLO cross section for the final state $F+$jet that can thus be computed with known NLO subtraction techniques such as, for instance, the dipole subtraction method\cite{Catani:1996vz}.
There are some extra singularities of pure NNLO type associated to the limit $q_T\to 0$, which are subtracted by the counterterm $d\sigma^{CT}_{NNLO}$.
If $F$ is colourless, $d\sigma^{CT}_{NNLO}$ is built by using the knowledge of the small-$q_T$ behaviour of the cross section as given by all-order resummation\cite{Bozzi:2005wk}.
The coefficient $\mathcal H^F_{NNLO}$ provides the contribution to the process with $q_T=0$, and thus contains the information on the virtual corrections.

The extension of eq.\ref{eq:qt_sub} to a massive colourful final state $F$ requires additional soft contributions, affecting both the counterterm $d\sigma^{CT}_{NNLO}$ and the coefficient ${\cal H}_{NNLO}^F$.
The contribution to $d\sigma^{CT}_{NNLO}$ is known from Ref.[\,\cite{Catani:2014qha}], while the contribution to ${\cal H}_{NNLO}^F$ has been recently evaluated by us \cite{inprep} (see also Ref.[\,\cite{Angeles-Martinez:2018mqh}]).

With the inclusion of those final ingredients, we were able to apply $q_T$-subtraction to the production of a colourful final state  and to carry out several phenomenological studies on $t\bar{t}$ production at NNLO.
We considered the inclusive cross section\cite{Catani:2019iny}, differential distributions and comparisons with experimental data\cite{Catani:2019hip}, and predictions in the $\overline{MS}$ renormalisation scheme for the top mass\cite{Catani:2020tko}.

Our calculation has been implemented in \textsc{Matrix}, a computational framework that allows the user to evaluate fully differential cross sections for a wide class of process at hadron colliders at NNLO QCD and NLO EW.
The core of \textsc{Matrix} is the Monte Carlo integrator \textsc{Munich}, that contains an implementation of the Catani-Seymour subtraction method\cite{Catani:1996vz,Catani:2002hc} for NLO QCD computations.
The required amplitudes are obtained either from \textsc{OpenLoops}\cite{Buccioni:2019sur} or, in the case of the two-loop amplitudes, in the form of numerical grids or analytic expressions.
\textsc{Matrix} then implements then the $q_T$-subtraction formalism and provides the user with the desired differential distribution for the chosen infrared-safe observable.

A preliminary version of the code for the computation of multi-differential distributions for $t\bar t$ production has been made available to experimental collaborations, and has already been used for data-theory comparisons\cite{CMS:2021vhb}.
In the latest release of \textsc{Matrix}, version 2.1.0, the inclusion of $t\bar t$ production has become publicly available\footnote{The latest version of \textsc{Matrix} can be downloaded at: \url{https://matrix.hepforge.org}}.

\textsc{Matrix} 2.1.0 contains several improvements with respect to the previous version.
The number of supported processes increases: not only $t\bar t$ production is included, but also tri-photon production\cite{Kallweit:2020gcp}.
Furthermore, the user has now access to the option to compute double-differential distributions for arbitrary infrared-safe observable.
Finally, on a more technical side, the inclusion of power corrections\cite{Buonocore:2021tke} and of the bin-wise extrapolation in the $r_{\rm cut}\to 0$ limit of the slicing parameter improve the accuracy of the predictions for the distributions.

\section{Further Developments: bottom-quark pair production}

\begin{figure}
\centerline{\includegraphics[width=0.65\linewidth]{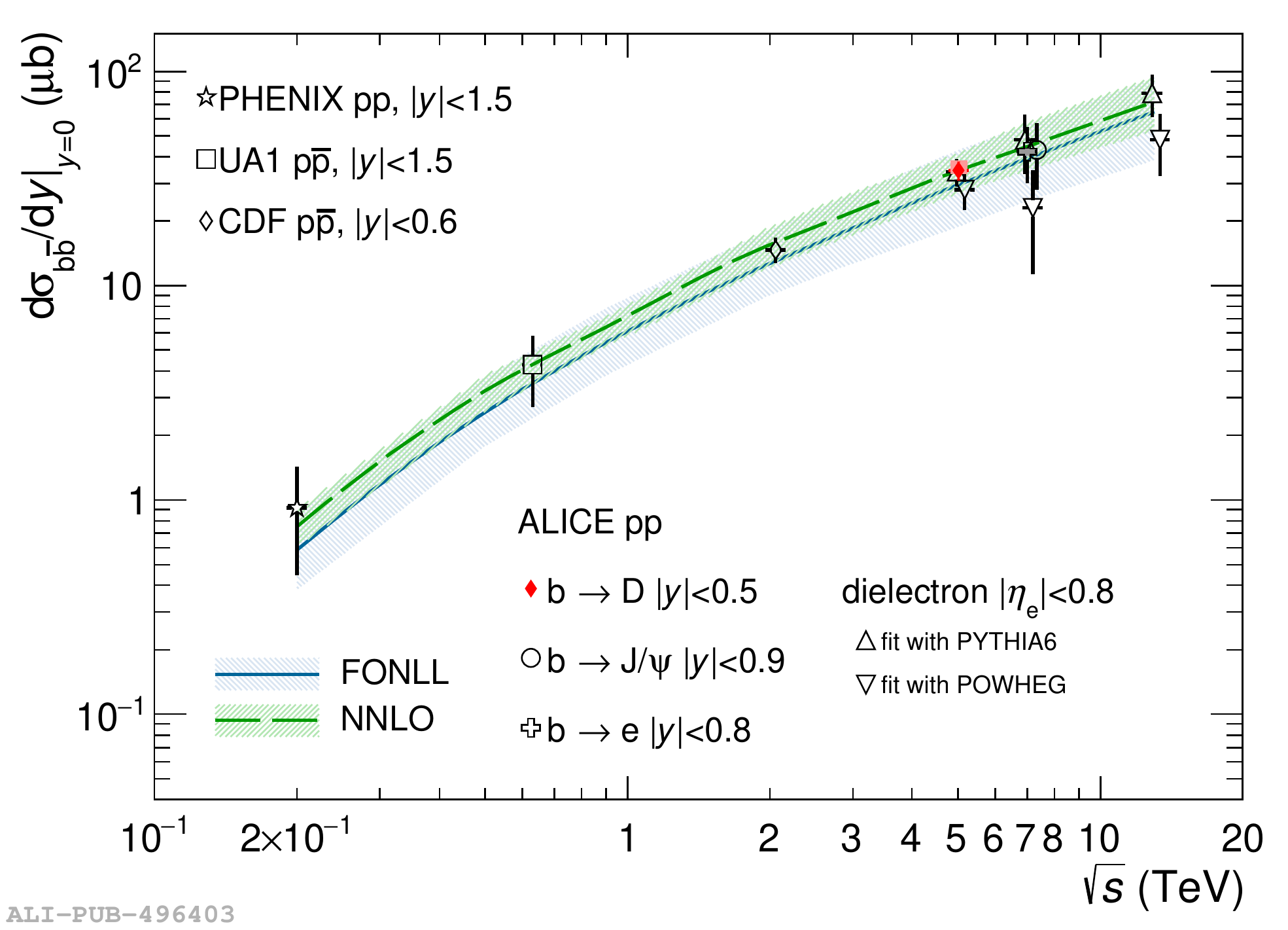}}
\caption[]{Bottom-quark pair production cross section as measured by the \textsc{Alice}, \textsc{Phenix}, CDF and UA1 collaborations. The green band represent the NNLO prediction computed with \textsc{Matrix}. Plot taken from Ref.[\,\cite{ALICE:2021mgk}].}
\label{fig:alice}
\end{figure}

The completion of the computation of $t\bar t$ production within the $q_T$-subtraction formalism opens the door to several possible extensions.

A first development was the computation of NNLO QCD corrections to $b\bar b$ production\cite{Catani:2020kkl}.
This has been implemented in the \textsc{Matrix} framework as an extension of $t\bar t$ production, by changing the heavy-quark mass and by allowing an arbitrary number of light flavours (from $n_f=5$ to $n_f=4$), which corresponds to a computation with a massive bottom quark and a top quark decoupled from the process.
We obtained results for the inclusive cross section
and first predictions for differential distributions.
The computation is fully implemented in a (at the moment) private release of \textsc{Matrix}. Predictions are provided upon request, and have already been used by the \textsc{Alice} collaboration in their analysis.

In Figure \ref{fig:alice} we show the comparison, presented by the \textsc{Alice} collaboration\cite{ALICE:2021mgk}, between different experimental measurements and theoretical predictions for $b\bar b$ production.
The \textsc{Alice} analysis corresponds to a centre of mass energy $\sqrt{s}=5.02$ TeV, and the cross section for $b\bar b$ production is extracted from D-meson measurements.
Figure \ref{fig:alice} additionally shows results at other collider energies as obtained by several other experiments (\textsc{Phenix}, CDF, UA1).
The data are compared to our NNLO predictions (green band) and to the FONLL predictions\cite{Cacciari:1998it} (blue band), which are obtained by combining NLO QCD corrections and resummed computations.
The inclusion of NNLO QCD corrections leads to a better agrement with the data at all the considered collider energies, and reduces the theoretical uncertainties, as obtained by using the customary 7-points scale variation.

\section{Conclusions and Outlook}
The $q_T$-subtraction formalism has been extended to the production of heavy-quark pairs in hadron collisions, leading to several phenomenological application to $t\bar t$ and $b\bar b$ production.
This computation has been implemented in the \textsc{Matrix} framework, and $t\bar t$ production is now supported as a process in the latest version of the public release of \textsc{Matrix}.

Our work can be extended in several directions. We can consider the production of a heavy-quark pair with additional colourless particles. First steps in this direction have been carried out with the computation of NNLO corrections for $t{\bar t}H$ production in the flavour off-diagonal channels \cite{Catani:2021cbl}. To complete the computation in the other channels some ingredients are still missing.
In particular, the two-loop amplitudes are not yet available, and the
contribution of soft final state emission has to be properly evaluated.

We can also consider the inclusion of top quark decays and off-shell effects. In this case, the process is actually $pp\to e^+\nu_eb\mu^-{\bar \nu}_\mu {\bar b}+X$ and higher-order QCD corrections can be computed by considering the production of the $b{\bar b}$ pair recoiling against the colourless $2l2\nu$ system. Work in these directions is ongoing.

\section*{References}
\bibliography{biblio}

\end{document}